\documentclass[11pt,a4paper,twoside]{article}
\usepackage{mathpazo}
\usepackage[a4paper,textwidth=13.5cm,top=3cm,bottom=3.2cm,twoside=false]{geometry}
\usepackage{xcolor}\definecolor{accent}{HTML}{000000}
\usepackage[utf8]{inputenc}
\usepackage[T1]{fontenc}
\usepackage[english]{babel}
\usepackage{amsmath,amssymb}
\usepackage{graphicx,booktabs}

\usepackage[font=small,labelfont=bf,labelsep=space]{caption}
\usepackage{csquotes}
\usepackage[hang,flushmargin]{footmisc}
\usepackage{microtype}
\usepackage{xcolor}
\usepackage{xurl}
\usepackage[colorlinks,allcolors=accent,hyperfootnotes=false]{hyperref}
\newenvironment{refs}{\section*{References}\small\begin{list}{}{\leftmargin1.5em\itemindent-1.5em\itemsep2pt\parsep0pt}\let\bibitem\item}{\end{list}}

\usepackage{titlesec}
\titleformat{\section}{\normalfont\large\scshape\color{accent}}{\thesection}{1em}{\MakeLowercase}
\titlespacing*{\section}{0pt}{2.2ex plus .5ex}{1.2ex}
\titleformat{\subsection}{\normalfont\itshape}{\thesubsection}{1em}{}
\titleformat{\subsubsection}{\normalfont\itshape\small}{\thesubsubsection}{1em}{}
\usepackage{fancyhdr}

\begin{document}
\thispagestyle{empty}
\begin{center}
\vspace*{1em}
{\large\bfseries Multi-field as a Determinable\par}
\vspace{1.4em}
{\scshape Davide Romano}%
{\renewcommand{\thefootnote}{\fnsymbol{footnote}}\footnote{University of Verona, Department of Human Sciences, Lungadige Porta Vittoria 17, 37129 Verona.\newline Email: \href{mailto:davide.romano@univr.it}{\nolinkurl{davide.romano@univr.it}}}\addtocounter{footnote}{-1}}\par
\vspace{1em}
{\color{accent}\rule{3cm}{0.4pt}}\par\vspace{0.6em}
{\small Published in C.~Ternullo \& M.~Antonelli (eds.),\\ \emph{Artificial Minds, Realism and Evidence in Science},\\ Isonomia -- Epistemologica XII (2025), pp.~245--275.\par}
\end{center}
\vspace{1.2em}
\begin{quote}\noindent{\scshape\color{accent} Abstract.}\enspace The paper advances the hypothesis that the \emph{multi-field} is a determinable, that is, a physical object characterized by indeterminate values with respect to some properties. The multi-field is a realist interpretation of the wave function in quantum mechanics, specifically it interprets the wave function as a new physical entity in three-dimensional space: a ``multi-field'' (Hubert \& Romano 2018; Romano 2021). The multi-field is similar to a field as it assigns determinate values to N-tuples of points, but is also different from a field as it does not assign pre-existing values at each point of three-dimensional space. In particular, the multi-field values corresponding to the empty points (points where no particles are located) have indeterminate values until a particle is located at those points. The paper suggests that the multi-field so defined can be precisely characterized in terms of determinable-based, object-level, account of metaphysical indeterminacy. Under this view, the multi-field as novel physical entity is, in fact, a metaphysically indeterminate quantum object, that is, a determinable.
\end{quote}
\vspace{0.8em}
\section{Introduction: the status of the wave function in quantum mechanics}

Quantum mechanics (QM) is a rather peculiar theory: on the one hand, it is a very successful theory and no one doubts that it grasps something \emph{true} about the nature of the microscopic world, on the other hand, however, the theory is metaphysically obscure, as the link between the formalism and the ontology of systems is not clear. As a result, different interpretations have proposed different ways to connect the formalism with the ontology. The standard interpretation of quantum mechanics (SQM, the theory presented in QM textbooks)\footnote{E.g. Landau-Lifshitz (2013); Sakurai \& Napolitano (2020); Shankar (2012).} is manifestly an operational theory, and in doing so it renounces to provide a realist description of systems. Literally taken, SQM indicates the spectrum of possible measurement results (eigenvalues) and their relative probability distribution. This theory works very well in practice, but it does not provide an ontology of systems independently from measurement. For example, a plane wave cannot be interpreted as physical wave, since the ontology of SQM concerns the eigenvalues and not the wave function per se and, most importantly, the wave function cannot be generally defined as a classical field in 3D space. This is consistent with the standard interpretation, where the wave function is a probability amplitude (not an ontological entity) and the ontology of the theory concerns the eigenvalues/measurement outcomes, but it leaves the question about the nature of quantum systems basically unanswered.\footnote{For an analysis of the ontology of standard quantum mechanics, see e.g. Ballentine (2014), Bowman (2008), Maudlin (2019), Norsen (2017).}

In order to overcome these limitations imposed by SQM, a certain number of non-standard interpretations—such as the Many Worlds Interpretation (MWI)\footnote{Wallace (2012).} and Relational Quantum Mechanics (RQM)\footnote{Rovelli (1996).}—as well as non-standard theories—such as the Ghirardi-Rimini-Weber (GRW)\footnote{Ghirardi \emph{et al. }(1986).} and the de Broglie-Bohm (dBB)\footnote{Bohm (1952); Bohm \& Hiley (1993).} theories—have been proposed in the last decades.\footnote{I distinguish between non-standard interpretations (MWI, RQM), which do not change the formalism of quantum mechanics (even though the MWI does not include the collapse postulate and may well be considered a non-standard theory as well) and non-standard theories, which do change the formalism of the theory by modifying the Schrödinger's equation (GRW theory) or the definition of the state (dBB theory).} All these approaches attempt to retrieve a realist description of quantum systems\footnote{The situation is actually more nuanced: the GRW and Everett theories have been originally proposed as solutions of the measurement problem and unification of the micro and macro regime, whereas the dBB theory has been proposed not to solve the measurement problem but to provide a realist account of quantum systems. Moreover, this theory was originally proposed by Louis de Broglie in 1927 during the Solvay conference, so it is historically as old as the Copenhagen interpretation (see e.g. Bacciagaluppi \& Valentini (2009) for the early history of the de Broglie’s theory).} while leaving the empirical predictions of SQM intact.\footnote{All non-standard theories must recover the empirical results of quantum experiments, which are successfully described by standard quantum mechanics.} It is a hard attempt, and in fact one that originated much controversy in the philosophical literature: controversy concerning the best theory to adopt, controversy on the correct metaphysical interpretation for each of the non-standard theories. Leaving aside the former issue, the latter one is closely connected to the interpretation of the wave function.

While SQM and RQM are clear on the status of the wave function, as in both theories the wave function is purely instrumental,\footnote{I list Relational Quantum Mechanics among the \emph{realist }interpretations because, even if this theory is instrumental on the wave function, it describes objective \emph{relative facts, }which are the result of any interaction between systems and do not depend on measurement operations. However, the debate on the metaphysics of RQM is still open in the literature and one may provide arguments to interpret RQM as an instrumental theory \emph{tout court}. While I would rather defend the former option on this point, the latter one is not excluded. } this question is genuinely open in the MWI, GRW and dBB theories. Since all of these theories aim to provide a realist account of quantum mechanics, the wave function also seems to take more than just an instrumental role. In particular, three different interpretations of the wave function in these theories have been proposed so far:
\begin{enumerate}
\item The nomological interpretation\footnote{Goldstein \& Zanghì (2013).}
\item The 3N-D field interpretation\footnote{Albert (2013); Ney (2021). I use the term ``3N-D field interpretation'' rather than ``wave function realism'' as there are different ways in which one can be realist on the wave function. For example, the multi-field approach is definitely a realist interpretation of the wave function, but it differs substantially from Albert’s and Ney’s wave function realism.}
\item The multi-field interpretation\footnote{Forrest (1988); Belot (2012); Hubert \& Romano (2018); Romano (2021a).}
\end{enumerate}
Even if it is common to discuss these approaches on similar grounds, it must be noticed that they have different areas of applicability, i.e. not all these interpretations can be consistently applied to all theories. In particular, the \emph{nomological view} and the \emph{multi-field approach} have a more restrictive application than the \emph{3N-D field interpretation}. While the latter can be applied to MWI, GRW and dBB theories, the nomological interpretation can be applied only to theories with a primitive ontology, that is, to dBB, $\mathrm{GRW}_m$ and $\mathrm{GRW}_f$\footnote{$\mathrm{GRW}_m$ and $\mathrm{GRW}_f$ stands, respectively, for ``GRW with mass-density'' and ``GRW with flashes''.} theories. The multi-field approach is even more restrictive, as it can be consistently applied only to the dBB theory (either in the first-order Bohmian mechanics or in the second-order Bohm’s 1952 theory).\footnote{The reasons for this restriction will be clear in sect.~3.3.}

It may be useful to summarize these remarks in the following table (1.1):

\begin{table}[h]
\centering
\small
\begin{tabular}{@{}p{3.6cm}p{8.4cm}@{}}
\toprule
\textbf{Interpretation of the wave function} & \textbf{Areas of applicability}\\
\midrule
3N-D field & MWI, BM (1\textsuperscript{st}-order), Bohm's theory (2\textsuperscript{nd}-order), $\mathrm{GRW}_0$, $\mathrm{GRW}_m$, $\mathrm{GRW}_f$\\
Nomological view & BM (1\textsuperscript{st}-order), $\mathrm{GRW}_m$, $\mathrm{GRW}_f$\\
Multi-field & BM (1\textsuperscript{st}-order), Bohm's theory (2\textsuperscript{nd}-order)\\
\bottomrule
\end{tabular}
\caption*{\textbf{Table 1.1}}
\end{table}

This paper does not want to enter in the debate concerning the best interpretation of the wave function, as (I am fairly convinced that) this is eventually left to personal preferences and perspectives. The aim of the present paper is instead more humble and, at the same time, more concrete: I want to provide a precise \emph{metaphysical characterization of the multi-field} in terms of the \emph{determinable-based account of metaphysical indeterminacy}.

The paper has the following structure: in (sect. 2) I review the interpretation of the wave function in standard QM; in (sect. 3) I describe the nomological view (sect 3.1), the 3N-D field interpretation (sect. 3.2) and the multi-field approach (sect 3.3). In (sect. 4) I present the determinable-based account and, following Wilson (2013, 2017), connect it to metaphysical indeterminacy. Finally, in (sect. 5) I describe the multi-field as a novel physical entity in terms of the determinable-determinate account. In (sect. 6) I draw some connections between the multi-field as determinable and relevant features of Bohm’s theory. Conclusions are given in (sect. 7).

\section{The wave function in standard quantum mechanics}

In SQM the wave function of a system represents completely the state of the system but it has no ontological significance: it is rather an instrument for computing (given a certain observable) the spectrum of possible eigenvalues and their probability distribution. The meaning of the \emph{state} is one of the most relevant differences between quantum and classical mechanics. In classical mechanics the system’s state (initial position and velocity) has a direct reference to the system’s ontology: the state represents the position and velocity of the system at the initial time. The classical state has therefore a double role, representational and ontological: it specifies the degrees of freedom needed to compute the evolution of the system (representational role) and, at the same time, it refers to a concrete system in space and time (ontological role). In SQM, instead, the state is just representational: the wave function represents the complete information on the system’s state, i.e. the information needed to compute the evolution of the system (via the Schrödinger’s equation), but it does not have a direct link with the system’s ontology: we do not know what kind of system the wave function represents, e.g. if the system \emph{is} a particle, or a wave, or a ``wave-particle'' or just a novel entity.

The physical meaning of the wave function in SQM is given by its absolute square $|\psi|^2$, which is interpreted as a probability density (Born’s statistical interpretation). Consequently, the integral of this quantity: $\int|\psi|^2\,dx$ gives the probability to obtain specific eigenvalues for specific observables. So construed, SQM is a theory about the possible measurement results (eigenvalues) of different observables. The eigenvalues are obtained from the collapse of the wave function, which is a postulate of the theory: in a measurement of the observable $A$, represented by the Hermitian operator $\hat{A}$, the wave function collapses instantaneously in one of the eigenstates $|a_k\rangle$ of $\hat{A}$ and the measurement result is mathematically represented by the eigenvalue $a_k$ associated to $|a_k\rangle$.\footnote{The Hermitian operator is defined by the action on its eigenstates: $\hat{A}|a_k\rangle=a_k|a_k\rangle$.}

We note that the eigenvalue cannot be assigned to the system before the collapse takes place, that is, before and independently of the measurement process. This means that a quantum superposition (e.g. a superposition of different locations in the two-slit experiment or a superposition of ``spin-up'' and ``spin-down'' in the singlet state of the electron) cannot be interpreted as a superposition of different eigenvalues as the collapse has not yet occurred.\footnote{The only exception being if the state is an eigenstate of a given observable, according to the eigenvalue-eigenstate link. In this case, the state does not change during the measurement process, so the collapse does not apply and we can assign a specific eigenvalue to the state independently from the measurement.} In such contexts, we must refrain to associate to the system a superposition of real-existing classically incompatible values since, according to the postulates of quantum mechanics, we can associate physical values to a quantum system only through eigenvalues and we cannot associate any eigenvalue before a measurement has taken place. We should distinguish instead between \emph{representational} and \emph{ontological }capacity of the wave function, where the former is the ability to mathematically represent the system and the latter the ability to indicate which kind of entity the system described by the wave function is. While SQM succeeds in the former task, it leaves the question about the ontology (latter task) basically unanswered. This aspect is unsatisfactory if we want to provide an ontology for quantum systems and is the main reason to look at the non-standard theories introduced above. Therefore, we now turn to the metaphysical analysis of the wave function in such theories.

\section{The wave function in non-standard quantum mechanics}

Differently from SQM, the wave function can take an ontological meaning in MWI, GRW and dBB theories.\footnote{I refer specifically to MWI since the original Everett’s theory (``relative state formulation'' of QM) is much closer in spirit to relational quantum mechanics: it does not postulate the existence of branching parallel worlds, it is observer-dependent and it describes sequences of ``records'' relative to the observers in line with RQM’s relative facts.} In these theories the spectrum of the metaphysical interpretations of the wave function is much greater: it can have an \emph{instrumental role} (as commonly assumed in the GRW physics community), or a \emph{nomological role} (as in the nomological view, where it guides the motion of the primitive ontology) or an \emph{ontological role}, where it can represent a physical entity (as in the 3N-D field interpretation and the multi-field approach) or just patterns in three-dimensional space (as in Wallace’s MWI).\footnote{Wallace (2010).} The debate is open and there is no consensus on what the wave function \emph{is }or \emph{represents }in these theories. In the next subsections I will present the three major interpretations that have been proposed so far in these contexts, namely the \emph{nomological view}, the \emph{3N-D field interpretation} and the \emph{multi-field} approach.

\subsection{The nomological interpretation}

The nomological interpretation has been originally proposed by Goldstein \& Zanghì (2013), even though some traces of this interpretation date back to Hiley and Bohm’s (1993) notion of \emph{active information}. According to this interpretation, the wave function in the dBB theory is a nomological entity, i.e. a mathematical object that has no ontological counterpart but that is necessary to describe the evolution of the system. The analogy is with the Hamiltonian function in classical mechanics: as the Hamiltonian function (mathematically represented in phase space) ``guides'' the motion of the particles in 3D space, the wave function (mathematically represented in configuration space) ``guides'' the motion of the Bohmian particles in 3D space via the guiding equation:\footnote{For simplicity, I write the guiding equation for spinless particles.}
\begin{equation}
\dot{q}=\frac{\hbar}{m}\,\mathrm{Im}\left(\frac{\nabla\psi}{\psi}\right)
\tag{3.1}
\end{equation}
That is: given $\psi(x,t)$ we can compute the velocity and so the trajectories of the Bohmian particles in the same manner as we can compute the trajectories of classical particles in Hamiltonian mechanics from $H(x,v)$. This interpretation is quite attractive as it dissolves the problems linked to the multi-dimensionality of the wave function (the fact that, for an N-particle system, the wave function is defined in 3N-D space rather than in 3D space), but it also faces important issues. For example, a typical nomological entity (like the classical Hamiltonian) is not time-dependent and is not contingent (i.e. it does not depend on the boundary conditions), while the wave function is contingent and (generally) time-dependent. In order to solve this problem, Goldstein and Zanghì posit that only the wave function of the universe--the \emph{Universal Wave Function} (UWF)—has an ontological significance. The UWF, as a solution of the Wheeler-de Witt equation, is supposed to be unique and time-independent.

However, this creates a further problem: in quantum mechanics we typically assign wave functions to (isolated) subsystems, never to the universe as a whole, exception made for quantum cosmology. That the UWF is the only wave function that counts from the ontological point of view is a metaphysical postulate. In addition, the very definition of UWF is not obvious: the wave function of the universe, if it exists, may well be represented by a factorized state between different (effective) wave functions, as it is plausible to assume that not all the regions and parts of the universe have previously interacted with each other, forming a unique entangled state. Furthermore, even leaving aside the problems associated to the universal wave function, there is a fundamental structural asymmetry between a nomological entity like the Hamiltonian, which has a \emph{bottom}-\emph{up }structure, i.e. it is built ``from below'' by the sum of the kinetic and potential energy of the particles, and a typical wave function, which has a \emph{top-down} structure, i.e. it is derived as a solution of a dynamical equation (the Schrödinger’s equation), as the electromagnetic field is a solution of the Maxwell’s equations.\footnote{See e.g. Romano (2021a, sect. 2).}

\subsubsection{Active information vs nomological view}

The idea of \emph{active information} proposed by Bohm \& Hiley (1987: 327-328) shares some common features with the nomological interpretation: according to Bohm \& Hiley, the wave function is a sort of information pool that guides the motion of the particles in the same manner as the electromagnetic waves produced by a remote guide the motion of a radio-controlled boat. As the electromagnetic waves transmit to the boat the information about its future movement (so the boat moves according to the transmitted information), the wave function transmits to the Bohmian particles the information on their future trajectory (so the Bohmian particles move according to the information transmitted by the wave function). The idea of active information may be seen as a precursor of the nomological interpretation, as it is a first attempt to regard the wave function as a non-material object (information pool, nomological entity) which guides the motion of a material object (Bohmian particles). The way it transmits this information to the particles is different in the two cases, but the general schema (action of a non-material entity to a material entity) is the same. Not really ``what the Doctor orders'', but what the information transmits.

\subsection{3N-D field interpretation}

The \emph{3N-D field interpretation}, most commonly known as \emph{wave function } \emph{realism}, has been originally proposed by Albert in two papers (1996, 2013). The idea is to interpret the wave function as close as possible to the role it plays in the quantum formalism: the wave function looks like a physical field, for it is the solution of a dynamical equation and, like a field, it assigns specific values to each point of the space on which it is defined. Since the wave function is defined on the system’s configuration space, it assigns values to each point of that space, not to points of three-dimensional space.

Based on these features of the quantum formalism, Albert proposes to regard the wave function as a \emph{physical field in configuration space.} As a classical field (e.g. the electromagnetic field) assigns real values to points of 3D space, the wave function assigns complex values to points of configuration space. According to this view, the wave function is analogous to a classical field, the only difference being that it is defined in configuration space rather than in three-dimensional space and that assigns complex values rather than real values to each point of its domain. There is however an important consequence: since the wave function is a physical field in configuration space, the latter must be recognized as the fundamental physical space of quantum mechanics. And since quantum mechanics is more fundamental than classical mechanics,\footnote{A fair consensus has been reached in the literature that environmental decoherence plays an important role in the quantum-to-classical transition (Joos \emph{et al.} (2013), Schlosshauer (2007, 2019); Zurek (2002)). According to this picture, the classical world emerges from the quantum world when quantum systems interact strongly and continuously with the external environment. The classical world in this picture is emergent or derivative with respect to the quantum world. See, however, Romano (2022) for a critical assessment of environmental decoherence in the standard context. A dissident voice on the importance of environmental decoherence for the classical limit is Ballentine (2008).} configuration space must be seen as the fundamental space of physics \emph{tout court}. This is why this position is better defined as \emph{configuration space realism} rather than wave function realism: one may be realist on the wave function without endorsing the fundamentality of configuration space (as it happens e.g. in the multi-field approach).\footnote{Furthermore, we note that configuration space, differently from three-dimensional space, does not have a fixed number of dimensions, as the latter depends on the number of degrees of freedom of the system under analysis. In particular, given an N-particle entangled state, the system’s configuration space has a number of dimensions $3N$, where $N$ is the number of particles composing the system. The number of dimensions therefore vary from system to system. This reflects the derivative character of configuration space with respect to three- dimensional space and, to my opinion, it is not a firm ground to assess the fundamentality of configuration space over three-dimensional space.}

If configuration space is fundamental, then the configuration of the Bohmian particles must be also placed on that space rather than on three-dimensional space. Following this reasoning, Albert reduces the (somewhat illusory or emergent) configuration of particles in 3D space to a ``marvelous point'' in 3N-D space. The marvelous point solves the communication problem\footnote{The problem of communication arises when the wave function and the Bohmian particles ``live'' on different spaces (e.g. Callender 2015) and can be summarized as follows: how does the wave function (object in 3N-D space) guide the Bohmian particles in 3D space?} between the wave function and the Bohmian particles as they are both placed in the same space, but it does not help with the ``perception problem'', i.e. the problem to understand why we perceive the macroscopic world as three-dimensional even though the fundamental space is the configuration space. Solutions to the perception problem have been proposed by Albert (2013) and more recently by Ney (2021). Albert argues that it is the structure of the Hamiltonian that decomposes configuration space into sets of three-dimensional coordinates, giving the impression that this is the space where particles move and interact with each other, and eventually giving the impression that we live in 3D space. Ney’s argument relies instead on the role of symmetries in quantum mechanics. In particular, she notes that, even though the fundamental space is configuration space, important symmetries of quantum mechanics are retrieved only when we represent systems in 3D space. In both cases, however, an open question remains on how these mathematical structures (Hamiltonian, symmetries) can affect our perception to live in a 3D world.

\subsection{The multi-field approach: the wave function as a new entity in 3D space}

In the multi-field view, the wave function is the mathematical representation of a \emph{multi-field}, which has to be regarded as a novel physical entity in 3D space. The idea of the multi-field comes originally from the notion of ``polywave'' proposed by Forrest (1988). Forrest interprets the wave function in SQM as a ``polywave'', that is, as a multiple assignment of field values for any ordered set of position coordinates. The notion of polywave has been then revisited and inserted in the context of Bohm’s theory by Belot (2012), who names it ``multi-field''. However, Belot dismisses quickly the multi-field idea, principally because of the non-validity of the action-reaction principle (while the multi-field acts on the Bohmian particles, the latter do not act back on the former).\footnote{A response to Belot on the back-reaction problem is given in Romano (2021a, sect 6.3).} After this first attempt, the multi-field approach has been further developed and defended in Hubert \& Romano (2018) and more recently in Romano (2021a).

As we saw before, the wave function looks like a field as it defines a specific value for each point of the space on which it is mathematically defined, yet these values are associated to points of configuration space and (for N-particle entangled states) they cannot be reduced to an assignment of pre-existing values associated to points of 3D space. In other words, the wave function generally assigns a continuous distribution of complex values to each point of the system’s configuration space. The idea of the multi-field is to interpret such distribution of values in configuration space as the mathematical representation of a novel physical entity in 3D space. More precisely, the assignment of definite values in configuration space is interpreted not as a classical field in configuration space, but rather as a novel kind of object in 3D space. The new object is the multi-field.

Even though the wave function cannot assign pre-existing, determinate values to each point of 3D space, a projection from configuration to three-dimensional space can always be done, as configuration space is literally the space of possible configurations of particles in three-dimensional space. We can illustrate this idea with the following example. Consider a system of two point-particles in 3D space represented by the coordinates $p_1(x_1,y_1,z_1)$ and $p_2(x_2,y_2,z_2)$: we can represent the 2-particle system as two discrete particles in 3D space or, equivalently, as a single particle in 3N-D space:

Two-particle system in configuration space
\begin{equation}
p_{12}(x_1,y_1,z_1,x_2,y_2,z_2)
\tag{3.3.1}
\end{equation}
We note, however, that the position coordinates of the two particles in the single particle representation are ordered: the first set of three coordinates $(x_1,y_1,z_1)$ represent particle $p_1$, the second set $(x_2,y_2,z_2)$ represent particle $p_2$. Configuration space is built from the configuration of particles in 3D space: we define $p_{12}$, a single point in 3N-D space, from the configuration of two particles$p_1$ and $p_2$ in 3D space. Since configuration space is derivative from three-dimensional space, the latter can be safely viewed as the fundamental physical space, i.e. the arena where systems exist and interact with each other. However, we are still left with the initial problem: the wave function assigns precise values to points of configuration space, not to points of 3D space. Let us clarify this point in the context of Bohm’s theory.

Consider a 2-particle entangled state. In Bohm’s theory this system is represented by a six-dimensional wave function $\psi_{12}(x_1,x_2,\dots,x_6)$ and by the actual particles’ configuration $q_{\mathrm{tot}}$ composed of two point-particles, $q_1$ and $q_2$, having exact locations and mathematically represented by the position coordinates in 3D space: $q_1(x_1,y_1,z_1)$ , $q_2(x_2,y_2,z_2)$. The Bohmian system is thus represented by the state: $(\psi(x),q_{\mathrm{tot}})$.

The wave function specifies a map from configuration space to the complex numbers:
\begin{equation}
\psi_{12}(x_1,x_2,x_3,x_4,x_5,x_6)\rightarrow c
\tag{3.3.2}
\end{equation}
If we leave the interpretation at this stage, we have the original idea of Forrest’s polywave (the wave function assigns a complex value to any ordered N-tuples of points), but in Bohm’s theory the wave function is always accompanied by the actual configuration $q$. When we insert the particles’ configuration into the wave function:
\begin{equation}
\psi(q_1,q_2)=\psi(x_1,y_1,z_1,x_2,y_2,z_2)
\tag{3.3.3}
\end{equation}
the wave function assigns a complex value to the two discrete points, $(x_1,y_1,z_1)$ and $(x_2,y_2,z_2)$, in 3D space corresponding to the exact location of the Bohmian particles $q_1$ and $q_2$. In general, for an N-particle system, the wave function assigns a complex value to the N-tuple of points corresponding to the actual particles’ configuration. The result is not a classical field, as it assigns simultaneously a specific complex value $(c)$ to an N-tuple of points in 3D space (in our case at the two points $(q_1,q_2)$) and the value is not pre-assigned but depends on the position coordinates of the particles composing the configuration $q$. Under this view, the wave function is thus a new kind of physical field--a multi-field--which assigns specific field values to N-tuples of points corresponding to the exact location of the Bohmian particles.

The multi-field so described can be thought as a generalization of a classical field: while a classical field (e.g. the electromagnetic field) assigns a determinate value to any point of 3D space, the multi-field assigns a determinate value only to N-tuple of points, corresponding in Bohm’s theory to the actual position of the particles. For example, given a wave function of the type:
\begin{equation}
\psi(x,y)=A\cos(xy)
\tag{3.3.4}
\end{equation}
and configuration $q=(q_1,q_2)$, with A a normalization constant, the multi-field assigns a determinate value:
\begin{equation}
\psi(q_1,q_2)=A\cos(q_1q_2)
\tag{3.3.5}
\end{equation}
in correspondence of the two points  $q_1$and $q_2$ occupied by the Bohmian particles. The determinate is computed by evaluating the 2-particle wave function at the points $x=q_1$ and $y=q_2$.

Note that the multi-field cannot be thought of as a continuous distribution of (determinate) values, differently from a classical field. While a classical field defines a determinate value at any point, the multi-field defines a determinate value only at those points where the Bohmian particles are located, leaving all the other (empty) points with indeterminate values. This constitutes a discontinuity in the field, and a primary difference with a classical field.\footnote{However, think this is not the kind of discontinuity that should be handled mathematically, since the multi-field represents a novel entity with respect to a classical field, one that is fundamentally characterized by discontinuity. This fundamental discontinuity is reflected in the metaphysical characterization of the multi-field as determinable (sect. 5).}

The multi-field assigns a determinate complex value to a given N-tuple of points at any instant, corresponding to the exact location of the Bohmian particles at that instant. If complex values may sound unphysical, we note that the wave function can be reduced to two (coupled) real-valued functions, corresponding to the amplitude $R(x,t)$ and phase $S(x,t)$ of the wave function written in polar form: $\psi(x,t)=R(x,t)\,e^{\frac{i}{\hbar}S(x,t)}$. Consequently, the complex-valued multi-field can be reduced to two (coupled) real-valued multi-fields associated to $R(x,t)$ and $S(x,t)$.\footnote{This approach has been proposed in Romano (2021a). Regarding the amplitude and phase of the wave function as multi-fields provides physical support to the quantum potential $Q$ and quantum force $F_Q$, which enters in the definition of the quantum Newton's law: $F_C+F_Q=m\ddot{q}$. In fact, the quantum force is generated by the quantum potential: $F_Q=-\nabla Q$ and the latter is generated by the amplitude of the wave function: $Q=-\frac{\hbar^2}{2m}\frac{\nabla^2R}{R}$.}

In practice we do not know the exact location of the Bohmian particles but we know that, given a system with wave function $\psi$, the actual configuration is statistically distributed according to the Born’s rule: $\rho(q)=|\psi(x)|^2$. This postulate is known as \emph{quantum equilibrium} and guarantees that the de Broglie-Bohm theory is empirically equivalent to standard QM. Since a Bohmian system is defined, at any time, by a unique actual configuration $q_{t^*}$, the multi-field assigns, at any time, a unique and specific value to the N-tuple of points ($x_1,\dots,x_N$) corresponding to $q_{t^*}=(q_1,q_2,\dots,q_N,t^*)$:
\begin{equation}
\psi(q_1,q_2,\dots,q_N,t^*)\rightarrow c
\tag{3.3.6}
\end{equation}
At any time, the multi-field assigns a complex value $c$ to the N-tuples of points in three-dimensional space: $q_1,q_2,\dots,q_N$, corresponding to the exact location of the Bohmian particles. Differently from Forrest’s polywave, in Bohm’s theory the multi-field assigns a unique determinate value associated to the configuration $q$ at any instant of time. Even if we do not know the precise location of the particles (but only that they are distributed according to the Born rule), this is as a matter of fact an epistemic ignorance and does not affect the ontology described so far: even if the actual configuration is epistemically unknown, still \emph{the Bohmian particles have an exact location in } \emph{3D space, }so the ontology of the multi-field is unambiguously determinate.

The multi-field can be regarded as a generalization of a classical field.\footnote{The multi-field as determinable is a novel physical entity with respect to a classical field in the same manner as the classical field is a novel physical entity with respect to the Newtonian force.} Whereas a classical field assigns a specific value to each point of 3D space, the multi-field assigns a non-local value to N-tuples of points of 3D space. I say ``non-local'' as the specific value assigned at one point (corresponding to the exact location of one particle of the configuration) depends non-locally (i.e. simultaneously at a distance) on the exact location of all the other particles of the configuration. It remains a problem, however: the multi-field does not specify any determinate value to the \emph{empty points}, i.e. all points in the domain of the wave function that are not occupied by the Bohmian particles. This raises a prompt objection: how do we know that the multi-field includes also the empty points (i.e. that is an entity distributed in space rather than a relation between N points) if the latter are not associated to any determinate value? In order to solve this problem, I propose the following hypothesis: the multi-field is a \emph{determinable} representing metaphysical, object-level, indeterminacy (QI). The multi-field represents an indeterminate state of affairs (Wilson 2013), yet it is ontologically as real as a classical field or a classical point-particle. Before expanding on this point, we introduce in the next section the determinable-determinate account and its link with metaphysical indeterminacy.

\section{Metaphysical indeterminacy}

Metaphysical indeterminacy (MI) is the idea that there is a state of affair of the world that is indeterminate and that such indeterminacy is intrinsic of the world itself. Such indeterminacy is therefore different from epistemic indeterminacy (coming from the lack of knowledge) or semantic indeterminacy (coming from vagueness or ambiguity of language). We may say: \emph{semantic indeterminacy} comes from an imperfect correlation between the language and a determinate world, \emph{epistemic indeterminacy} comes from an incomplete knowledge of the determinate world, \emph{metaphysical } \emph{indeterminacy} is the acknowledgment that the world itself is indeterminate. Metaphysical indeterminacy divides into two main accounts: \emph{meta-level} and \emph{object-level} MI (Wilson 2013). The former is represented by \emph{metaphysical } \emph{supervaluationism (}Barnes 2010; Barnes \& Williamson 2011); the latter by the \emph{determinable-based} or \emph{determinable-determinate }account (Wilson 2013, 2017). Following Wilson (2013) and Calosi \& Mariani (2021), the difference between these two accounts of metaphysical indeterminacy is that:

\begin{quote}
[A]ccording to the former [metaphysical supervaluationism] it is indeterminate which determinate state of affairs obtains (SOA), whereas according to the latter [determinable-based account] it is determinate that an indeterminate SOA obtains. [Calosi \& Mariani (2021: 8)]
\end{quote}

\subsection{Supervaluationism}

\emph{Metaphysical supervaluationism} can be roughly summarized by the following quote by Barnes (2010: 622):

\begin{quote}
It’s perfectly determinate that everything is precise, but [...] it’s indeterminate which precise way things are.
\end{quote}

Calosi \& Mariani (2021: 9) describe how supervaluationism can be applied to quantum mechanics, in particular how a superposition state can be described using supervaluationism:

\begin{quote}
In general, consider a system S in state $|\omega\rangle=c_1|\psi\rangle+c_2|\phi\rangle$. There is MI because there are two admissible precisifications, the SOA that $\psi$ and that $\phi$ respectively, and it is indeterminate which one is the case. That is, superposition indeterminacy boils down to indeterminacy about which term of the superposition obtains.
\end{quote}

However, we can safely dismiss supervaluationism from our analysis for two reasons. First, supervaluationism does not seem to capture the characteristics of quantum mechanics. A superposition state is not a state in which the two eigenstates (precisifications, in this case) are determinate but we do not know which one obtains. This description does not capture the essence of a quantum superposition, in which all eigenstates (in the case above: $|\psi\rangle$ and $|\phi\rangle$) concur to the description of the behavior of the system, represented by the state vector $|\omega\rangle$ with different probability associated to each state (given by the absolute square of the associated coefficient). A state of affairs in which all the eigenstates of a superposition are equally determinate, as proposed by supervaluationism, would fail to generate the typical quantum interference that we observe in quantum experiments. For example, in the double-slit experiment with electrons or photons, the interference pattern that is progressively generated on the screen can be accounted for only considering constructive and destructive interferences between the two components between the slits and the screen, and the latter can be accounted for only considering different amplitudes between the interfering components. This tension is reported in Calosi \& Mariani (2021: footnote 17): ``we should note that the straightforward application raises questions on how to understand the coefficient $c_1$ and $c_2$ in the quantum state''.

Furthermore, supervaluationism so defined seems to collapse into epistemic indeterminacy. If the world is totally precise and composed of multiple determinates, then it is just a matter of convention or lack of knowledge which one of these determinates represents the actual world. For example, this is how Darby (2010: 235) applies metaphysical supervaluationism to the Schrödinger’s cat paradox:\footnote{The quote is reported in Calosi \& Mariani (2021: 8-9).}

\begin{quote}
[There is] a suggestive parallel between the terms in the superposition and the idea [...] of precisifications. One of the terms in the superposition [...] is a term where the cat is alive, the other is not; that is reminiscent of multiple ways of drawing the extension of ‘alive’, on some of which ‘the cat is alive’ comes out true, on some, false.
\end{quote}

We see that this description does not seem to capture the essential features of the paradox: the cat in the box (before a measurement is performed) is in a quantum superposition of being alive and dead, as the cat is in an entangled state with the radioactive material in the box, which is represented by a coherent superposition of two definite states, being decayed and not decayed (more precisely, the radioactive material is represented, in general, by a decreasing exponential function that describes the probability amplitude of the radioactive decay as a function of time). According to SQM the result of a measurement on the state of the cat will describe a determinate state of affair, but such determinate SOA cannot be ascribed to the eigenstates associated to the cat in the box before a measurement is performed. If supervaluationism does that, then it would be in conflict with standard quantum mechanics. A more promising approach is the determinable-based account introduced in the next section.

\subsection{Determinable-based account}

The\emph{ determinable-based account} of MI or \emph{determinable-determinate} account has been introduced by Wilson (2013, 2017) and later applied to quantum indeterminacy (e.g. Wolff (2015); Calosi \& Wilson (2018); Calosi \& Mariani (2021); Fletcher \& Taylor (2024)). The basic idea is that a state of affair is described by a property or an object represented by a \emph{determinable }and a \emph{determinate}, the two standing in a specific property-type relation. The determinable is more general and accounts for a spectrum of possible determinates, the determinate is a specific instance or realization or actualization of the determinable. This is, for example, how Wilson (2017) presents the determinable-determinate account:

\begin{quote}
Determinables and determinates are in the first instance type-level properties that stand in a distinctive specification relation: the ``determinable–determinate'' relation (for short, ``determination''). For example, color is a determinable having red, blue, and other specific shades of color as determinates; shape is a determinable having rectangular, oval, and other specific (including many irregular) shapes as determinates; mass is a determinable having specific mass values as determinates.
\end{quote}

We can report the cited examples of determinable-determinate relations in the following table (4.1):

\begin{table}[h]
\centering
\small
\begin{tabular}{@{}ll@{}}
\toprule
\textbf{Determinables} & \textbf{Determinates}\\
\midrule
Color & Red, blue, green, \dots\\
Shape & Rectangular, oval, \dots\\
Mass & Mass values $m_1$, $m_2$, \dots\\
\bottomrule
\end{tabular}
\caption*{\textbf{Table 4.1}}
\end{table}

As reported in the quote above by Wilson, a standard example of determinable-determinate relation concerns the property of \emph{color. }By saying that an object is ``colored'' we specify a \emph{determinable}: a property (the property of being colored) to which may correspond many specific instances (the spectrum of determinate colors). If we say that a certain (colored) object is ``red'' we specify a determinate (a specific, determinate color) for the given determinable (being colored). The determinable account is pyramidal: ``red'' is a determinate with respect to the determinable ``being colored'' but is a determinable with respect to different shades of red, such as ``scarlet'' or ``vermillion''.

Note that in all these examples the determinable does not exist independently from the determinate: it does not exist in the world a colored object without a specific color, or a shaped object without a determinate (regular or irregular) shape. We anticipate that the multi-field is a determinable of a different type: it is a determinable object which exists independently of its determinate. This is valid for the multi-field account presented here as well as for any application of the determinable-based account to quantum indeterminacy (e.g. Calosi \& Wilson (2018)).

In physics, the determinable-based account has been applied to classical properties such as \emph{mass} (of a classical system) and to quantum properties such as the \emph{position }(Bokulich (2014)) and \emph{spin }(Wolff (2015)) of a quantum system. There is however an important difference between the classical and quantum case. In the classical case, the determinable property is always accompanied by a determinate. Consider, for example, the mass of a table. We may say that the mass as determinable is the general property of a classical object (a table in this case) of having a mass. However, it does not exist a classical object that has a mass without having a specific mass value. That is, in classical physics, the determinable (e.g. the mass property) is always accompanied by a determinate (a specific mass value). Same for colors or shapes: it does not exist a colored object without a specific color, or a shaped object without a specific shape.

This is not the kind of relation between determinable and determinate that we find in quantum mechanics. A quantum system that is in a superposition of eigenstates with respect to a certain observable does not have a specific value for that observable (before a measurement is performed). The observable in quantum mechanics can thus be represented by a determinable without a determinate. Two standard examples concern the position and the spin of a quantum system. Consider a 1-particle system represented by a plane wave:
\begin{equation}
\psi(x)=A\,e^{\frac{i}{\hbar}px}
\tag{4.2.1}
\end{equation}
where $A$ is a normalization constant and $p$ the momentum eigenvalue. This state indicates an equal probability distribution to find the particle in any point of the space in a position measurement:
\begin{equation}
P_x=|\psi(x)|^2=|A|^2
\tag{4.2.2}
\end{equation}
Until a measurement is performed, the particle does not have an exact position in space, that is, the observable ``position'' has a determinable without a determinate. The example of plane wave is summarized by Bokulich (2014: 467) as follows:

\begin{quote}
In quantum theory it is more typically the case that the degree to which the particle’s momentum is specified allows us to say, for example, that the particles is located somewhere in this room, although it is not possible to say that is located in any particular point in the room. In other words, while it makes sense to talk about the particle having the property of position (that is to say the particles are in the room), that property cannot be ascribed a definite (precise) value.
\end{quote}

To be precise, in standard quantum mechanics we cannot say that the particle ``is located somewhere'' before the measurement is performed, as this would imply an epistemic interpretation of quantum probabilities, which is in conflict with the standard interpretation. It would be more correct to say that the particle is located \emph{nowhere} before the measurement. Consequently, in the determinable-based account of MI, a quantum system (in SQM) never has a determinate position if not in the precise instant of a position measurement.\footnote{The situation is even more tricky: position eigenstates are represented in SQM by Dirac delta functions, which are not solutions of the Schrödinger’s equation. In practice, a quantum system is considered fairly localized in position when it is represented by a Gaussian or a well-localized state.}

The example of spin as determinable is analyzed by Wolff (2015). The spin case is different from the position case as the latter is a scalar quantity while the former is a vectorial quantity. For this reason, the spin operator is always defined along a given direction, so we have three different operators: $\hat{S}_x,\hat{S}_y,\hat{S}_z$, which represent the spin operator, respectively, along the $x-$, $y-$ and $z-$ axis. Consider a $\frac{1}{2}-$spin particle (e.g. an electron): this particle has two possible eigenvalues or the spin $\left(+\tfrac{1}{2};-\tfrac{1}{2}\right)$, respectively associated to the eigenstates ``spin-up'' $|\uparrow\rangle$ and ``spin-down'' $|\downarrow\rangle$. As the three operators $\hat{S}_x,\hat{S}_y,\hat{S}_z$ are mutually incompatible (so it does not exist a state that is an eigenstate simultaneously of two of these operators), when the electron has a determinate spin along a given direction, the spin along a different direction is represented by a superposition of two eigenstates and thus is not determinate.

From this analysis Wolff suggests that we must associate a determinable to each individual operator $\hat{S}_x,\hat{S}_y,\hat{S}_z$ and not to the spin property \emph{tout }court. Furthermore, Wolff notes that while the determinable-based account describes well the relation between the spin property and the spin value along a given direction, it does not explain why the operators $\hat{S}_x,\hat{S}_y,\hat{S}_z$ are mutually incompatible, i.e. it does not explain why certain sets of determinables cannot have joint determinates (the same conclusion applies to all sets of non-commuting observables, such as e.g. position and momentum).

Finally, Wolff analyzes three approaches to correlate the spin as a determinable with metaphysical indeterminacy. The first is the one proposed by Funkhouser (2006: 566), according to which: ``an amendment for the quantum level might be that every object instantiating a determinable also instantiates certain determinates to certain probabilities.''\footnote{The quote is reported in Wolff (2015, p. 383).} This approach however does not work in the case of spin: the determinate (``spin up'' or ``spin down'') is always a well-defined value, while probabilities are associated to uncertainty about the specific measurement result, as reported by Wolff (2015: 384):

\begin{quote}
[w]hat exactly the probabilities denote is of course controversial, but minimally they simply state the likelihood of finding a particle with spin value ``up'' and ``down'' respectively in a given direction. By adding in the probabilities, we simply seem to acknowledge the indeterminacy of the spin state, we don’t eliminate it.
\end{quote}

The second and third approaches are instead those proposed by Wilson (2013): we can think of a determinable as corresponding to the instantiation of multiple determinates (``glutty'' MI) or to the instantiation of none of the determinates (``gappy'' MI). In the first case, we should think of the different directions of the spin as different but complementary perspectives. The classical example is the iridescent feather where multiple determinates colors are realized with respect to different perspectives. In the case of the electron spin:

\begin{quote}
[T]his would mean that we treat the determinate outcomes of spin measurement in different directions as different perspectives. Depending on which measurement we carry out, i.e. how we orient our Stern-Gerlach device, we will get a determinate z-spin up, say, or a determinate y-spin down, but it would be misleading to suggest that the electron only has a determinate z-spin or only a determinate y-spin. It is just that from the perspective (read: measurement) we have chosen, this is the determinate which is realized in our perspective. [Wolff (2015: 384)]
\end{quote}

This approach also encounters a number of convincing objections. First, it looks very closely to an epistemic reading of quantum uncertainty, furthermore there is a difference between multiple determinates of the same determinable (e.g. spin up and spin down along $x-$direction) and multiple determinates associated to different determinables (e.g. spin up and spin down along the $y-$direction for the state $|\uparrow\rangle_z$) that does not seem to be correctly described by this approach. Building on this analysis, Wolff concludes (convincingly, in my opinion) that the approach considering ``gappy'' MI is the best one of the three:

\begin{quote}
Of the three answers to the question of indeterminacy, then, the third seems to be the most promising. It is also the most radical revision of the determinables/determinate distinction, since it requires the instantiation of determinables without determinates. If that is to be possible, determinables have to be accepted into the ontology on equal footing with determinates. [Wolff 2015, p. 385]\footnote{Wolff also reports some reservation on this kind of approach, as it requires a radical revision of the current ontology: ``It is not obvious that this is a price worth paying, given how little the application of the determinables model seems to contribute to our understanding of quantum indeterminacy.'' (Wolff 2015, p. 385).}
\end{quote}

This is very close to the idea proposed here and developed in the next section to characterize the multi-field as a novel physical object. This idea imposes a radical revision of the current ontology, but one that (likely) offers more clarity in the interpretation of quantum indeterminacy and, in general, in the interpretation of the quantum ontology. We note that applying the determinable-based account to the multi-field is a step further with respect to applying it to spin or position in SQM, as the multi-field is not a property of the system but (part of) the system itself in the de Broglie-Bohm theory. Under this novel approach, the determinable does not describe the properties of a system but the system itself: the wave function is interpreted as a multi-field and the system, represented by the wave function (and by the particles’ configuration) is itself interpreted as a determinable, that is, as a new kind of object. In the next section we will expand on this point and characterize more precisely the multi-field as a determinable.

\section{The multi-field as a determinable}

The hypothesis presented here is that the wave function is the mathematical representation of a new physical entity, a \emph{multi-field},\footnote{The name ``multi-field'' is correct insofar we intend it as a true generalization of a classical field. This generalization is fully captured by the determinable-determinate account and provides an example of quantum indeterminacy. We note that the quantum indeterminacy introduced by the multi-field characterizes the entity itself, not the properties of the system. A Bohmian system has a definite position (specified by the actual configuration), a precise velocity (specified by the guiding equation), a precise acceleration (specified by the quantum Newton’s law), yet the multi-field values at the empty points have an indeterminate value.} which can be metaphysically characterized as a determinable, i.e. an object defined by properties without a determinate value. The multi-field is actually more complex than the determinable usually presented in the literature, as it assigns a determinate (a specific and unique complex value) to the N-tuple of points corresponding to the actual configuration of the Bohmian particles ($x_i=q_i$) and a determinable without a determinate to all the other points $x_i\neq q_i$. Following the determinable-based account of MI, the multi-field so defined implies ontological indeterminacy, i.e. it describes an indeterminate state of affairs in the world:

\begin{quote}
Here I present an account on which what it is for there to be MI is for it to be determinate (or just plain true) that an indeterminate (imprecise) SOA obtains. I more specifically suggest that the obtaining of an indeterminate SOA is profitably understood in terms of an object’s having, on the one hand, a determinable property, but not having, on the other hand, a unique property that is a determinate of that determinable. [Wilson (2013: 360-361)]
\end{quote}

Within the region $R$ where the multi-field is well-defined (the projection of the wave function in 3D space), the determinable property is represented by the (complex) values that the multi-field assigns to each point of three-dimensional space. It is a determinable as (i) the value of each of these points ($x_i\neq q_i$) is not determinate but, at the same time, (ii) a determinate is selected for any of these points once a particle is located at that point, i.e. when the initially empty point is included in the points corresponding to the actual configuration $x_i=q_i$. In other words, any empty point is characterized by a \emph{set of possible (potentially infinite) multi-field values.} A specific value from this set is selected, however, when a particle of the configuration $q$ is located at that point: the (originally empty) point will be so associated with a determinate, unique multi-field value.

This criterion of selection of the determinate is for some aspects similar to the way we select a value for a classical field, but for other aspects very different. Consider an electric field $\vec{E}(x,t)$ defined in the region $\Gamma$. This field assigns a specific value to any point $x\in\Gamma$ for any instant of time. The way we generally define a field value is associated to the indirect effect of the field on a charged test particle. For example, if we locate a test particle on the point $x_k\in\Gamma$ at time $t=t^*$, the particle will accelerate under the Lorentz force: $\vec{F}(x_k)=q\vec{E}(x_k,t^*)$. From the acceleration of the test particle we derive indirectly the existence of the electric field $\vec{E}(x,t)$ in that region. In the case of the multi-field we do not have test particles but we can divide the scheme between the wave function $\psi(x,t)$ and the Bohmian particles $q=(q_1,\dots,q_N)$. For simplicity, consider a two-particle state with wave function $\psi(x_1,x_2,t)$ and actual particle configuration $q=(q_1,q_2)$, defined in a one-dimensional potential box with length $L$. The points where the multi-field as determinable is well-defined correspond to the points where the wave function in 3D space is well defined, i.e. to all points: $0\leq x\leq L$. Differently from the electric field, the multi-field does not assign a specific value to each point of the region $0\leq x\leq L$, excluding the points $(x_1=q_1;x_2=q_2)$. Suppose, however, that we want to know the value of the multi-field associated to the (originally empty) point $x=\tfrac{L}{2}$. In this case, analogously to the case of the test particle, we can derive the (determinate) value of the multi-field at $x=\tfrac{L}{2}$ at the time $t=t^*$ by assuming to locate (as, in practice, we cannot control the position of Bohmian particles) one of the two particles of the configuration $q=(q_1,q_2)$ exactly at the point $x=\tfrac{L}{2}$. Suppose that we choose particle 1, represented by $q_1$: we thus consider the system $\psi(x_1,x_2,t^*)$ with particle configuration $q=\left(\tfrac{L}{2},q_2\right)$. In this case, the multi-field will assign the (complex) determinate value:
\begin{equation}
\psi\left(\tfrac{L}{2},q_2,t^*\right)=c
\tag{5.1}
\end{equation}
to the couple of points $\left(\tfrac{L}{2},q_2\right)$, that is, to the two points corresponding to the exact location of the Bohmian particles. We note from this example that the value of the multi-field at $q_1=\tfrac{L}{2}$ is determinate but \emph{non-local}, as it depends on the specific location $q_2$ of the other particle of the actual configuration.

The analogy here is that, as the test particle proves (indirectly) the existence of the electric field by the effect of the field on the particle, in a similar manner the effect on the Bohmian particle (the velocity via guiding equation or the acceleration via quantum Newton’s law) proves (indirectly) the existence of the multi-field. In particular, we can compute the determinate multi-field value at any point of the region where the multi-field is well-defined by locating (hypothetically) a Bohmian particle of the actual configuration at that point. This process transforms a determinable (a set of infinite possible values) into a determinate (a specific complex value). There are, of course, two important differences in the classical and quantum case. First, the Bohmian particle is not a test particle. While in the case of the electric field we assume to put an external particle (test particle) to evaluate the value of the field, in the case of the multi-field we assume to put a particle of the actual configuration that composes the Bohmian system. Second, as mentioned before, the value of the multi-field at the point $x=\tfrac{L}{2}$ depends non-locally on the value of $q_2$, i.e. the position of particle 2. At any instant $q_2$ will be represented by a specific real number, and overall the multi-field will assign a unique determinate to the couple of points $\left(\tfrac{L}{2},q_2\right)$. Yet, if we change the location of the second particle $q_2$ the multi-field value at $q_1=\tfrac{L}{2}$ will also changes, as the multi-field assigns one specific value for the entire configuration: $\psi\left(\tfrac{L}{2},q_2\right)=c$. Differently from the classical case, the determinate value of the multi-field at one point depends on the exact location of distant particles of the actual configuration. We may say that, differently from the classical case, the multi-field assigns a \emph{non-local} determinate value to the N-tuple of points corresponding to the actual configuration of particles: $(x_1=q_1,\dots,x_i=q_i,\dots,x_N=q_N)$. In this way, Bohmian non-locality (and quantum non-locality more generally) is implemented in the very definition of the multi-field. The multi-field as determinable can be naturally regarded as a \emph{non-local beable}.\footnote{On this point see also Hubert \& Romano (2018, sect. 5).}

The multi-field so defined is (plainly) a determinable: it describes an indeterminate but objective, ontologically real, state of affairs. This is exactly the state of affair associated to a determinable, as reported by Wilson (2013: p. 366):

\begin{quote}
\textbf{Determinable-based MI}: What it is for a state of affairs to be MI in a given respect R at a time t is for the state of affairs to constitutively involve an object (more generally, entity) O such that (i) O has a determinable property P at t, and (ii) for some level L of determination of P, O does not have a unique level-L determinate of P at t.
\end{quote}

In the \emph{multi-field-as-determinable} account, the MI state of affair involves the object or entity ``multi-field'' $M$ such that (i) $M$ has a determinable property $P$ at $t$, i.e. the multi-field values that it assigns at any empty point (excluding the points  $x_i=q_i$) within the region where the wave function in 3D is well-defined and (ii) for any point $x_i\neq q_i$, $M$ does not have a unique determinate of $P$ at $t$. There are two levels $L$ of determination: $L_1,L_2$. The first corresponds to the empty points within the multi-field region: $L_1(x_i\neq q_i)$, the second to the points of the actual configuration $L_2(x_i=q_i)$ For the level of determination $L_1$ there is no unique determinate of $P$: any point is associated with a set of possible multi-field values. For the level of determination $L_2$ there is instead a unique determinate: a specific complex value assigned to the N-tuple of points corresponding to the actual configuration $(x_1=q_1,\dots,x_N=q_N)$.

The metaphysical indeterminacy implied by the determinable-based account can be characterized even more precisely. In fact, there are two ways in which a determinable can fail to have a unique determinate: either it has none, or it has more than one. The former case is termed ``gappy'' MI, the latter ``glutty'' MI. A standard definition is given in Calosi (2021: 11305):

\begin{quote}
According to the Determinable Based Account (DBA) of metaphysical indeterminacy (MI), there is MI when there is an indeterminate state of affairs, roughly a state of affairs in which a constituent object x has a determinable property but fails to have a unique determinate of that determinable. There are different ways in which x might have a determinable but no unique determinate: x has no determinate—gappy MI, or x has more than one determinate—glutty MI.
\end{quote}

The multi-field as determinable is a case of \emph{gappy metaphysical } \emph{indeterminacy}, as the determinable $P$ fails to assign a determinate value at any point $x_i\neq q_i$. In conclusion, the multi-field as determinable is defined as a distribution of determinable-property $P$, that is, a set of possible complex values for each point within the region of 3D where the wave function is well-defined. At any empty point ($x_i\neq q_i$) corresponds a determinable without a determinate, however the point takes a determinate as soon as it is occupied by a particle ($x_i=q_i$). The specific value at that point will depend not only on the wave function but also on the exact location of distant particles that compose the actual configuration, so defining a \emph{non-local determinate}.

\section{Some remarks on the ontology of the multi-field and Bohm’s theory}

In this final section, I present some remarks on the metaphysics of the multi-field as determinable in connection with relevant features of Bohm’s theory, in particular with the nature of non-locality, the guiding equation and the quantum equilibrium. These remarks are not intended to be complete, but they want to offer a suggestion on the metaphysical import of the multi-field view within the ontology of Bohm’s theory.\footnote{Thanks to an anonymous reviewer for inviting me to clarify these points concerning Bohmian non-locality, the guiding equation and the Born’s probabilistic distribution in the multi-field-as-determinable view.}

\subsection{Multi-field as determinable and non-locality}

From the discussion above, we notice that the multi-field as determinable implements Bohmian (and in general quantum) non-locality quite naturally, as the determinate depends at the same time on the precise location of all the Bohmian particles. Changing the position of one particle of the configuration instantaneously changes the determinate value that the multi-field assigns at that configuration. As suggested above, we can say that the determinate is \emph{non-local}, according to this description. Consequently, the multi-field as determinable view accounts for the non-local correlations between distant particles (for N-particle entangled states) since the determinate value of the multi-fields depends instantaneously on the exact position of all the Bohmian particles of the configuration, no matter how distant they are. The Bohmian particles follow the actual trajectories guided by the guiding equation, but even when these particles are at space-like distance, the determinate value of the multi-field at a given time will depend on the exact location of the particles at that time. This is the way in which the multi-field accommodates the experimental violation of Bell’s inequalities: the determinate cannot be locally defined, its value will be defined at any instant only by the actual configuration of the Bohmian particles, independently from the distance between the particles. For example, given a 2-particle entangled state:
\begin{equation}
\psi(x_1,x_2)=c_1\psi_1(x_1)\psi_2(x_2)+c_2\psi_2(x_1)\psi_1(x_2)
\tag{6.1.1}
\end{equation}
with actual configuration $q=(q_1,q_2)$, when the entangled state describes a macroscopic superposition, e.g. when the two components  $\psi_1(x_1)\psi_2(x_2)$ and $\psi_2(x_1)\psi_1(x_2)$ are at a macroscopic distance with each other (this is also the case of space-like separated components) the Bohmian particles $(q_1,q_2)$ will enter only one of the two components, giving rise to the effective factorization.\footnote{The process of effective factorization or \emph{effective collapse }has been originally introduced in Bohm \& Hiley (1987). In short, the effective factorization is the process that originates \emph{effective wave functions} from larger entangled states when the latter describe macroscopic superpositions. This is the Bohmian equivalent of the branching process in Many Worlds Interpretation. Note that the formation of effective wave functions (EWFs) is independent from the interaction with the measuring apparatus. For example: in Bohm’s theory, the entanglement between the system and the external environment produces EWFs (see e.g. Romano 2023). The formation of EWFs is the physical basis of decoherence in Bohm’s theory.} As a result, we have two possible cases:

\begin{align}
&1.\quad \psi_1(q_1)\psi_2(q_2)\quad\text{with probability } P=|c_1|^2 \tag{6.1.2}\\
&2.\quad \psi_2(q_1)\psi_1(q_2)\quad\text{with probability } P=|c_2|^2 \tag{6.1.3}
\end{align}
Repeating the experiment several times, this will result in the usual non-local correlations described by Bell’s theorem. Note that every time the multi-field will have a determinate value described by $\psi_1(q_1)\psi_2(q_2)$ or $\psi_2(q_1)\psi_1(q_2)$.

\subsection{Multi-field and the guiding equation}

It must be noticed that, even though the multi-field assigns indeterminate values to most points of the wave function, the velocity of the Bohmian particles, described by the guiding equation, is defined for the N-tuples of points corresponding to the actual location of the Bohmian particles. And for these points the multi-field assigns a determinate. For the empty points (corresponding to indeterminate values of the multi-field) the guiding equation can still be defined, but it does not correspond to a real velocity of the particles. In other words, the guiding equation defines a velocity field for all points of the wave function, but the actual velocity of the particles is defined only for the points occupied by the particles. For these points the multi-field has a determinate. This grounds an ontological correspondence between the multi-field as determinate and the real velocity of the particles. The particles’ velocity is always defined at their actual location, and the actual location of the particles correspond to the N-tuple of points for which the multi-field assigns a \emph{determinate}.

\subsection{Determinate and indeterminate knowledge}

From the ontological point of view, the multi-field assigns a unique determinate at any instant. The determinate is assigned at the N-tuple of points where the Bohmian particles are located. However, from the epistemic point of view, the exact position of the Bohmian particles is unknown and statistically distributed according to: $\rho(q)=|\psi(q,t)|^2$. Consequently, the maximum knowledge we can have of the determinate value of the multi-field will be also statistically distributed according to the Born’s rule. The fact that we do not know epistemically the exact configuration at a given instant, however, is not relevant for the ontology of the multi-field: independently from our knowledge, the state of affair (metaphysically speaking) is determinate: there is a unique location of the particles at every instant, which corresponds to a unique determinate of the multi-field and many (potentially infinite) indeterminate values for the unoccupied points. To this regard, the multi-field does not pretend to explain why the Bohmian particles are statistically distributed according to the Born’s rule, or why this statistical distribution represents an ultimate epistemic constraint. This is an assumption that we have to maintain in the multi-field account, as it happens in all other metaphysical interpretations of the wave function in Bohm’s theory, such as the nomological and the realist interpretation in configuration space.

\section{Conclusions}

The multi-field can be characterized as a determinable, as it assigns to each point of 3D space a set of possible, potentially infinite, complex values and a determinate to the N-tuples of points which correspond to the exact location of the Bohmian particles. The multi-field so defined is a case of ``gappy'' metaphysical indeterminacy: it describes an indeterminate state of affairs in which a determinable property is instantiated by a set of possible determinates. The determinate specified by the multi-field is non-local, as it depends from the actual position of the Bohmian particles composing the system. Under this approach, the pilot-wave of the de Broglie--Bohm’s theory becomes an object less concrete and more abstract than a classical wave, but one that guides physically the particles in 3D space.

\section*{Acknowledgments}
I wish to thank the audience at the Triennial International Conference of the \emph{Italian Society for Logic and the Philosophy of Science} in Urbino 2023, where this paper was originally presented. This work has been supported by the \emph{Fundação para a Ciência e a Tecnologia} through the fellowship FCT Junior Researcher hosted by the Centre of Philosophy of the University of Lisbon.

\begin{refs}\bibitem Albert, D. Z. (2013), ``Wave function realism'', in D. Albert \& A. Ney (eds.), \emph{The Wave Function: Essays on the Metaphysics of Quantum Mechanics}, Oxford, Oxford University Press.
\bibitem Bacciagaluppi, G. \& Valentini, A. (2009), \emph{Quantum Theory at the Crossroads: Reconsidering the 1927 Solvay Conference}, Cambridge University Press.
\bibitem Ballentine, L. E. (2008), ``Classicality without decoherence: a reply to Schlosshauer'', in \emph{Foundations of Physics}, n. 38, pp. 916--922.
\bibitem Ballentine, L. E. (2014), \emph{Quantum Mechanics: A Modern Development}, World Scientific Publishing.
\bibitem Barnes, E. (2010), ``Ontic vagueness: a guide for the perplexed'', in \emph{Noûs}, vol. 44, n. 4, pp. 601--627.
\bibitem Barnes, E. \& Williams, R. (2011), ``A theory of metaphysical indeterminacy'', in K. Bennett \& D. W. Zimmerman (eds.), \emph{Oxford Studies in Metaphysics}, vol. 6, Oxford, Oxford University, pp. 103--148.
\bibitem Belot, G. (2012), ``Quantum states for primitive ontologists: a case study'', in \emph{European Journal for Philosophy of Science}, n. 2, pp. 67--83.
\bibitem Bohm, D. (1952), ``A suggested interpretation of the quantum theory in terms of ``hidden'' variables'', part I and II, in \emph{Physical Review}, vol. 85, n. 2, pp. 166–193.
\bibitem Bohm, D. \& Hiley, B. J. (1987), ``An ontological basis for the quantum theory: non-relativistic particle systems'', in \emph{Physics Reports} vol. 144, n. 6, pp. 321--375.
\bibitem Bohm, D. \& Hiley, B. J. (1993), \emph{The Undivided Universe: An Ontological Interpretation of Quantum Theory}, Routledge.
\bibitem Bokulich, A. (2014), ``Metaphysical indeterminacy, properties, and quantum theory'', in \emph{Res Philosophica}, vol. 91, n. 3, pp. 449--475.
\bibitem Bowman, G. E. (2008), \emph{Essential Quantum Mechanics}, Oxford, Oxford University Press.
\bibitem Callender, C. (2015), ``One world, one beable'', in \emph{Synthese}, vol. 192, n. 10, pp. 3153--3177.
\bibitem Calosi, C. \& Mariani, C. (2021), ``Quantum indeterminacy'', in \emph{Philosophy Compass}, vol. 16, n. 4, e12731.
\bibitem Calosi, C. \& Wilson, J. (2018), ``Quantum metaphysical indeterminacy'', in \emph{Philosophical Studies}, pp. 176, 1--29.
\bibitem Calosi, C. (2021), ``Gappy, glutty, glappy'', in \emph{Synthese}, vol. 199, n. 3, pp. 11305--11321.
\bibitem Fletcher, S. C. \& Taylor, D. E. (2024), ``The representation and determinable structure of quantum properties'', in \emph{Synthese}, vol. 204, n. 44.
\bibitem Forrest, P. (1988), \emph{Quantum Metaphysics}, Blackwell.
\bibitem Ghirardi, G. C., Rimini, A. \& Weber, T. (1986), ``Unified dynamics for microscopic and macroscopic systems'', in \emph{Physical review D}, vol. 34, n. 2: 470.
\bibitem Goldstein, S. \& Zanghì, N. (2013), ``Reality and role of the wave function in quantum theory'', in D. Albert \& A. Ney (eds.), \emph{The Wave Function: Essays on the Metaphysics of Quantum Mechanics}, Oxford, Oxford University Press.
\bibitem Hubert, M. \& Romano, D. (2018), ``The wave function as a multi-field'', in \emph{European Journal for the Philosophy of Science}, vol. 8, pp. 521--537.
\bibitem Joos, E. Zeh, H. D., Kiefer, C., Giulini, D. J., Kupsch, J. \& Stamatescu, I. O. (2013), \emph{Decoherence and the Appearance of a Classical World in Quantum Theory}, Berlin, Springer.
\bibitem Landau, L. D. \& Lifshitz, E. M. (2013), \emph{Quantum Mechanics: Non-Relativistic Theory}, vol. 3, Elsevier.
\bibitem Maudlin, T. (2019), \emph{Philosophy of Physics: Quantum Theory}, Princeton, Princeton University Press.
\bibitem Ney, A. (2021), \emph{The World in the Wave Function: A Metaphysics for Quantum Physics}, Oxford, Oxford University Press.
\bibitem Norsen, T. (2017), \emph{Foundations of Quantum Mechanics}, Berlin, Springer.
\bibitem Romano, D. (2021a), ``Multi-field and Bohm’s theory'', in \emph{Synthese}, vol. 198, n. 11, pp. 10587--10609.
\bibitem Romano, D. (2021b), ``On the alleged extra-structures of quantum mechanics'', in \emph{Foundations of Physics}, vol. 51, n. 1, pp. 1--19.
\bibitem Romano, D. (2022), ``The unreasonable effectiveness of decoherence'', in V. Allori (ed.): \emph{Quantum Mechanics and Fundamentality}, \emph{Naturalizing Quantum Theory between Scientific Realism and Ontological Indeterminacy}, vol. 460, Springer, pp. 3--18.
\bibitem Romano, D. (2023), ``A decoherence-based approach to the classical limit in Bohm’s theory'', in \emph{Foundations of Physics}, vol. 53, n. 41, part of the special issue by A. Drezet (ed.): \emph{Pilot-wave and beyond: Louis de Broglie and David Bohm's quest for a quantum ontology}, Foundations of Physics, 2023.
\bibitem Rovelli, C. (1996), ``Relational quantum mechanics'', in \emph{International Journal of Theoretical Physics}, vol. 35, pp. 1637--1678.
\bibitem Sakurai, J. J. \& Napolitano, J. (2020), \emph{Modern Quantum Mechanics}, Cambridge, Cambridge University Press.
\bibitem Schlosshauer, M. A. (2007), \emph{Decoherence and the Quantum-to-Classical Transition}, Berlin, Springer.
\bibitem Schlosshauer, M. A. (2019), ``Quantum decoherence'', in \emph{Physics Reports}, vol. 831, pp. 1--57.
\bibitem Shankar, R. (2012), \emph{Principles of Quantum Mechanics}, Springer Science \& Business Media.
\bibitem Wallace, D. (2010), ``Decoherence and ontology, or: how I learned to stop worrying and love FAPP'', in S. Saunders, J. Barrett, A. Kent \& D. Wallace (eds.), \emph{Many Worlds? Everett, Quantum Theory and Reality}, Oxford, Oxford University Press.
\bibitem Wallace, D. (2012), \emph{The Emergent Multiverse: Quantum Theory According to the Everett Interpretation}, Oxford, Oxford University Press.
\bibitem Wilson, J. (2013), ``A determinable-based account of metaphysical indeterminacy'', in \emph{Inquiry}, vol. 56, n. 4, pp. 359--385.
\bibitem Wilson, J. (2017), ``Determinables and determinates'', \emph{Stanford Encyclopedia of Philosophy} (first published: 2017; substantive revision: 2023). Online: \url{https://plato.stanford.edu/archives/spr2023/entries/determinate-determinables/}
\bibitem Zurek, W. H. (2002), ``Decoherence and the transition from quantum to classical - revisited'', in \emph{Los Alamos Science}, vol. 27, pp. 86--109.
\end{refs}
\end{document}